\documentclass[aps,prl,showpacs,twocolumn]{revtex4}
\usepackage{graphicx} 
\usepackage{dcolumn} 
\usepackage{multirow}
\begin{document} 

\title{Identification of the Au coverage and structure of the Au/Si(111)-5$\times$2 surface}
\author{Se Gab Kwon}
\author{Myung Ho Kang}
\affiliation{Department of Physics, Pohang University of Science and Technology, Pohang 790-784, Korea}
\date{\today}

\begin{abstract}

We identify the atomic structure of the Au/Si(111)-5$\times$2 surface by using density functional theory calculations.  
With seven Au atoms per unit cell, our model forms a bona fide 5$\times$2 atomic structure, which is energetically favored over the leading model of Erwin, Barke, and Himpsel [Phys. Rev. B {\bf 80}, 155409 (2009)] and well reproduces the Y- and V-shaped 5$\times$2 STM images.
This surface is metallic with a prominent half-filled band of surface states, mostly localized around the Au-chain area.
The correct identification of the atomic and band structure of the clean surface further clarifies the adsorption structure of Si adatoms and the physical origin of the intriguing metal-to-insulator transition driven by Si adatoms.

\end{abstract}

\maketitle


The Au/Si(111)-5$\times$2 surface is a representative self-organized one-dimensional (1D) metal chain system \cite{coll95,losi00,mcch04,choi08} and has served as a rich source of intriguing 1D phenomena such as atomic-scale Schottky barriers \cite{yoon04}, 1D domain-wall hoppings \cite{kang08}, and confined doping on a metallic chain \cite{bark12}.
Its atomic structure, however, is not yet solved in spite of extensive experimental studies \cite{bish69, bask90, scha91, maho92, mark95, hase96, benn02, kira03, yoon05, bark09, mcal10, kaut14} and density functional theory (DFT) calculations \cite{kang03,erwi03,riik05,renn07,chua08,erwi09}.
This long-standing surface science problem is thus considered a touchstone of our ability to look into the structure of complex surface/nano systems at truly atomic scale.


In particular, the structural debate was recently reignited by two conflicting reports \cite{abuk13,hoga13}.
Figure 1(a) shows the best structural model so far, proposed by a combined theoretical-experimental study of Erwin, Barke, and Himpsel (hereafter, EBH) in 2009 \cite{erwi09}.  
It features a Si honeycomb chain and an Au triple chain and contains six Au atoms and twelve top-layer Si atoms in a 5$\times$2 unit cell, well reflecting the experimental estimations of 5.6--6.7 Au atoms \cite{bark09,mcal10,kaut14} and 11--14 Si atoms \cite{tani90,chin07}.  
The EBH model was recently challenged by a new model derived by Abukawa and Nishigaya (hereafter, AN) in a reflection high-energy electron diffraction (RHEED) study \cite{abuk13}.
The AN model has the same Au and Si coverages as the EBH model but quite distinct structural features: 
It has no Si honeycomb chain, and the Au chain contains four Au rows rather than three. 
More recently, however, the AN model was excluded by Hogan $et$ $al$. \cite{hoga13}: The AN model was energetically and microscopically unfavored in DFT calculations, and their analysis of previous reflectance anisotropy spectroscopy data \cite{mcal10} strongly supported the presence of Si honeycomb chains, thereby being in favor of the EBH model. 


One inherent problem with the EBH model is that it is basically a 5$\times$1 structural model as seen in Fig. 1(a) and therefore is incompatible with the observed, distinct 5$\times$2 STM images \cite{maho92,benn02,kira03,yoon05}.
EBH argued by DFT calculations \cite{erwi09} that 5$\times$2 STM images may be possible from the 5$\times$1 structure with the aid of either Si adatoms or electron doping, but the origin of such electron doping is not clarified yet.


In this Letter, we report a new structural model for the Au/Si(111)-5$\times$2 surface. 
The key feature of our model is the incorporation of one more Au atom on the EBH model (that is, seven Au atoms per 5$\times$2 cell), which is found to resolve the inherent problems with the 5$\times$1 EBH model. 
In what follows, DFT calculations demonstrate that the new model is more energetically favored than the EBH model, reproduces the 5$\times$2 STM images, and well explains the Si-adatom related structural and electronic properties.  


We perform DFT calculations using the Vienna $ab$-$initio$ simulation package \cite{kres96} within the generalized gradient approximation \cite{perd96} and the projector augmented wave method \cite{bloc94, kres99}.
The Si(111) surface is modeled by a periodic slab geometry with eight atomic layers and a vacuum spacing of about 11 {\AA}.
The calculated value 2.372 {\AA} is used as the bulk Si-Si bond length. 
Au atoms are adsorbed on the top of the slab, and the bottom of the slab is passivated by H atoms.
We expand the electronic wave functions in a plane-wave basis with an energy cutoff of 250 eV.
A 2$\times$8$\times$1 $k$-point mesh is used for the 5$\times$2 Brillouin-zone integrations.
All atoms but the bottom two Si layers are relaxed until the residual force components are within 0.02 eV/{\AA}.
We confirmed for the proposed Au/Si(111)-5$\times$2 model that the formation energy and interatomic distances converge well within 0.01 eV and 0.02 {\AA}, respectively, by the used parameters.
We energetically compare different models by estimating the relative formation energy by $\Delta$$E=E_{\rm 1}$$-$$E_{\rm 0}$$-$$\Delta$$n_{\rm Au}$$\mu_{\rm Au}$$-$$\Delta$$n_{\rm Si}$$\mu_{\rm Si}$, where $E_{\rm 1}$ ($E_{\rm 0}$) is the total energy of a particular model (a reference model), $\Delta$$n$ is the number difference of the specified atoms relative to the reference model, and $\mu$ is the calculated bulk chemical potential for the specified atom.


\begin{figure*}
\centering{\includegraphics[width=16cm]{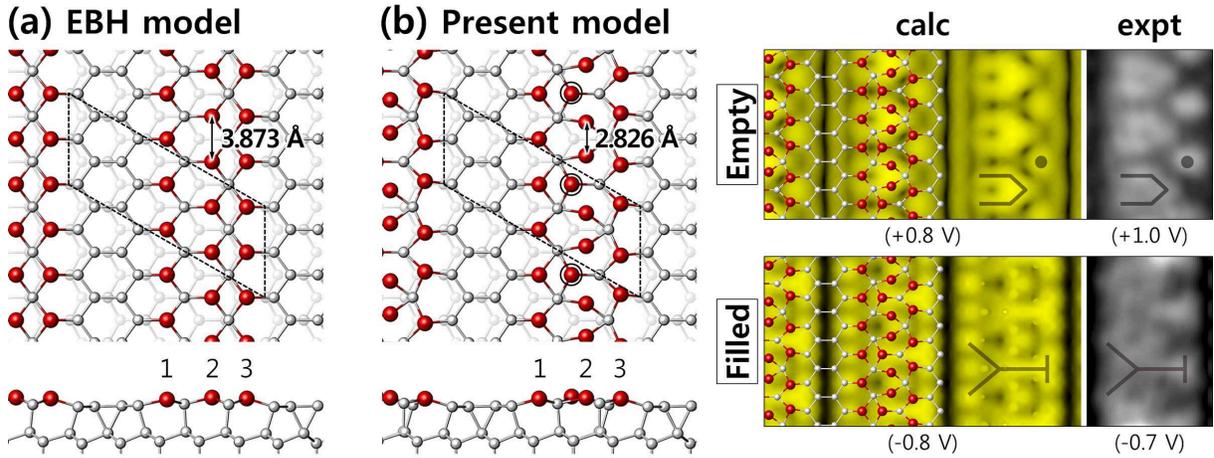}}
\caption{\label{fig1}
(Color online)
Structural models for Au/Si(111)-5$\times$2: (a) EBH model and (b) present model.
Large (small) balls represent Au (Si) atoms, and dashed lines represent a 5$\times$2 unit cell.
Three rows of Au atoms are denoted by numbers and the added Au atoms by circles in (b).
The simulated STM images represent the surface of constant density with $\rho$=2$\times$10$^{-6}$ e/\AA$^{3}$ taken at bias voltages +0.8 V (empty) and $-$0.8 V (filled). 
The round, V-shaped, and Y-shaped features are marked for easy comparison.
The experimental images were taken from Ref. \cite{erwi09}.
}
\end{figure*}


The basic idea of the present work is to add one Au atom per 5$\times$2 cell on the EBH model.
The incorporation of one more Au atom is not only a simple way of transforming into a desirable 5$\times$2 structure but also could possibly be a physical realization of the idea of electron doping given by EBH.
The resulting seven Au atoms per 5$\times$2 cell, which is referred to as 0.7 ML, is slightly out of the experimental estimations of 0.56--0.67 ML \cite{bark09,mcal10,kaut14} but is worth examining in light that a higher calibration of 0.65--0.67 ML is the latest result of a low-energy electron diffraction and microscopy study \cite{kaut14}. 


Figure 1(b) shows the lowest-energy structure where the added Au atom prefers to adsorb on a hollow site between the 1st and 2nd Au rows with at least 0.62 eV more gain in adsorption energy than other sites. 
More importantly, this new model has a much lower formation energy by 0.92 eV per 5$\times$2 cell than the EBH model. 
We find that, due to the added Au atom, the preexisting Au atoms undergo an intriguing 5$\times$2 reconstruction: While the 1st Au row is almost intact, the Au atoms in the 2nd and 3rd rows show substantial $\times$2 modulations along the chain direction.
Especially noticeable is the dimerization of Au atoms in the 2nd row with a bond length of 2.826 {\AA}, which reminds that the presence of a short Au dimer of 2.84 {\AA} was strongly suggested in the RHEED study of Abukawa and Nishigaya  \cite{abuk13}.


Indeed, the present model remarkably well reproduces the 5$\times$2 STM images of the clean surface.
As seen in Fig. 1(b), the simulations clearly resolve the well known V-shaped (empty-state) and Y-shaped (filled-state) STM features \cite{maho92, yoon05, erwi09}.
It is really encouraging that we are able to explain the 5$\times$2 STM images without the aid of either Si adatoms or electron doping \cite{erwi09}: In our model, a necessary 5$\times$2 reconstruction was preempted by the extra Au atom.  


\begin{figure}[!b]
\centering{\includegraphics[width=8cm]{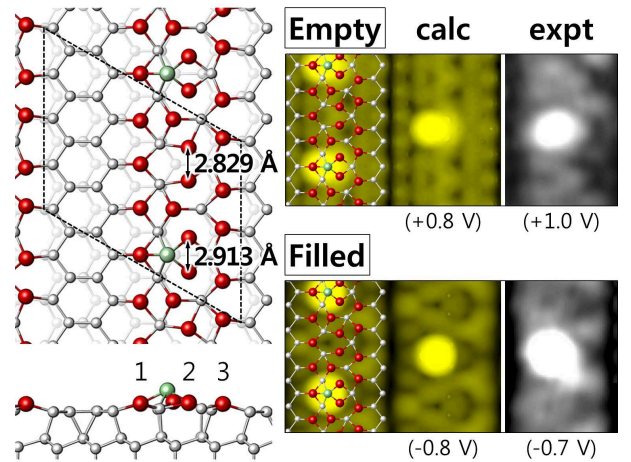}}
\caption{\label{fig2}
(Color online)
The Au/Si(111)-5$\times$4 surface with 0.05 ML Si adatoms.
STM images were taken in the same way as in Fig. 1.
}
\end{figure}


It is known that Si adatoms are always present on the Au/Si(111)-5$\times$2 surface, randomly occupying a unique adsorption site with at least $\times$4 spacing along the Au chain direction \cite{benn02,kira03}: 
The coverage of Si adatoms varies between 0.025 ML (the equilibrium coverage found in typical growth conditions) and 0.05 ML (the saturation coverage obtained by extra Si deposition), where 0.1 ML refers to one adatom per 5$\times$2 cell.
We first consider the saturated surface with 0.05 ML that is known to form a well-ordered 5$\times$4 Si-adatom phase  \cite{benn02,kira03}.
Figure 2 shows the lowest-energy structure where the Si adatom prefers to adsorb on a hollow site between the 1st and 2nd Au rows with at least 0.23 eV more gain in adsorption energy than other sites. 
We find that there is no significant change in the underlying 5$\times$2 surface structure: The bond length of the adjacent Au dimer, possibly the most affected by the Si adatom, increases marginally from 2.826 to 2.913 {\AA}, but the distant Au dimer is intact (2.829 \AA).
The simulated STM images also show that the effect of the Si adatoms is rather localized in a narrow range around themselves: Apart from their own $\times$4 bright protrusions, the 5$\times$2 substrate images are intact, well preserving the V- and Y-shaped features of the clean surface. 
The bright protrusion is located on a V-shaped feature (empty-state) and in between two Y-shaped features (filled-state), which are found to agree well with the experimental STM images \cite{maho92,yoon05,erwi09}.


\begin{table}
\caption{\label{tab:table1}
Adsorption properties of Si adatoms as a function of the coverage. 
$\Delta$$E$ (eV) is the relative formation energy per 5$\times$2 unit cell.
$h_{\rm Si}$ (\AA) is the average Si-adatom height from the Au layer. 
$d_{\rm Au}$ (\AA) represents the minimum and maximum lengths of the Au dimers.
}
\begin{ruledtabular}
\renewcommand{\arraystretch}{1.3}
\begin{tabular}{cccc}
Coverage & $\Delta$$E$ & $h_{\rm Si}$ & $d_{\rm Au}$ \\
\hline
0 ML & 0.000 & - & 2.826--2.826 \\
0.025 ML & 0.025 & 1.108 & 2.828--2.925 \\
0.05 ML & 0.073 & 1.107 & 2.829--2.913 \\
0.075 ML & 0.273 & 1.106 & 2.813--2.909 \\
0.1 ML & 0.461 & 1.106 & 2.878--2.878 \\
\end{tabular}
\end{ruledtabular}
\end{table}


\begin{figure*}
\centering{ \includegraphics[width=16cm]{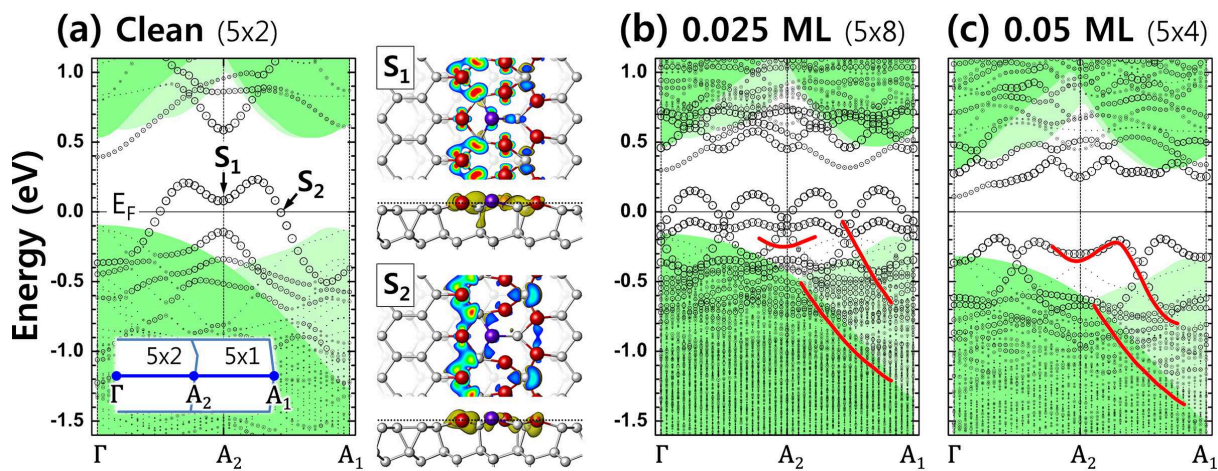}}
\caption{\label{fig3}
(Color online)
Band structure of the (a) clean, (b) 0.025 ML Si-adatom, and (c) 0.05 ML Si-adatom surfaces. 
The size of each circle is proportional to the contribution from the surface atoms in the Au-chain area.
Shaded areas represent the bulk band structure projected onto the 5$\times$2 surface (the 5$\times$1 projection is stressed by darker color).
In the charge plots, the top view represents the charge density in the horizontal plane marked in the side view.
The solid lines in (b) and (c) represent the ARPES data (Ref. \cite{choi08}) obtained for 0.024 ML and 0.048 ML, respectively.
}
\end{figure*}


Table 1 shows the properties of Si adatoms at different coverages: 
Here, we used a 5$\times$8 supercell and considered uniformly distributed Si-adatom phases for 0.025, 0.05, and 0.1 ML's (corresponding to $\times$8, $\times$4, and $\times$2 adatom lattices, respectively) and a nonuniform distribution for 0.075 ML.  
We find that the Si-adatom height and the Au-dimer lengths are rather insensitive to the coverage, but the formation energy relative to the clean surface increases rapidly with coverage.
While the energy differences are marginal by 0.025 eV per 5$\times$2 cell at 0.025 ML and by 0.073 eV at 0.05 ML, the values at higher coverages become substantial by 0.273 eV at 0.075 ML and 0.461 eV at 0.1 ML.
The deduced thermodynamic instability of the higher-coverage phases would be a good energetical account for the experimental saturation coverage of 0.05 ML \cite{benn02,kira03}.
For low coverages up to 0.05 ML, however, the energy differences are not sufficiently large enough for a conclusive thermodynamic discussion.
Experimentally, the saturated $\times$4 adatom phase at 0.05 ML is metastable and reverts by annealing to the equilibrium 0.025 ML coverage \cite{benn02}, and the equilibrium phase itself is not uniform but consists of $\times$4 adatom chains and empty segments in between \cite{kira03}, unlike the $\times$8 uniform adatom lattice we considered. 
Such nonuniform adatom distributions are beyond the scope of the present 5$\times$8 supercell approach.


Figure 3(a) shows the band structure of the clean Au/Si(111)-5$\times$2 surface.
This surface is metallic with a prominent half-filled band, in agreement with an earlier scanning-tunneling-spectroscopy  report that the adatom-free region is metallic \cite{yoon04}.
The band states are mostly distributed around the Au-chain area as shown in the charge characters of two representative states S$_{1}$ and S$_{2}$: While S$_{1}$ is rather localized at around individual surface atoms, S$_{2}$ appears as a typical 1D metallic state, delocalized along the 1st Au row.


The metallic band of the clean surface undergoes an interesting change by Si adatoms. 
Figure 3(b) shows the band structure modified by 0.025 ML Si adatoms (one adatom per 5$\times$8 cell).
Here, note that the original half-filled band of the 5$\times$2 surface becomes fourfold due to the use of a 5$\times$8 supercell: Two of the four half-filled bands are completely filled by the Si adatoms, and the other two still remain half filled.
This indicates that one Si adatom donates effectively two electrons to the Au chain. 
We found in our charge analysis that the states in the filled bands are rather localized around the Si adatom and the states in the half-filled bands mostly in the adatom-free region.
Doubling the Si-adatom density to 0.05 ML drives the remaining metallic bands completely filled, leading to an insulating band structure as seen in Fig. 3(c).
Therefore, 0.05 ML is the very coverage that completes the metal-to-insulator transition driven by Si adatoms, which provides an important physical meaning to the experimental saturation coverage.


In their angle-resolved photoemission spectroscopy (ARPES) study, Choi $et$ $al$. \cite{choi08} also observed that the band structure of this surface is converted from a metal to an insulator when the Si-adatom coverage is systematically increased in a range of 0.018--0.048 ML. 
Their ARPES data obtained for 0.024 ML and 0.048 ML are shown in Fig. 3.
In Fig. 3(c), the upper ARPES band is successfully reproduced by our calculation (again, the more calculated bands reflect the effect of the 5$\times$4 zone folding), which reinforces the validity of the underlying structural model.
The lower ARPES band, however, is missing in the calculation that shows no strong enough surface feature in that energy range.
A bulk origin may be suspected for this ARPES band, based on the earlier ARPES report by McChesney $et$ $al$. \cite{mcch04} that this band reveals a 5$\times$1 periodicity and our finding that it indeed disperses along the edge of the calculated 5$\times$1 bulk projection.
Fig. 3(b) shows a good agreement between theory and experiment even at 0.025 ML, but it should be stressed that the calculation was based on a $\times$8 uniform adatom lattice while the experiment on a nonuniform adatom phase \cite{kira03}.


In summary, we identified the Au coverage and structure of the Au/Si(111)-5$\times$2 surface through the energetical, microscopic, and spectroscopic examinations by DFT calculations.
Based on the correct atomic and band structure of the clean surface, we also verified the adsorption structure of Si adatoms and the physical origin of the intriguing metal-to-insulator transition driven by Si adatoms.


This work was supported by the National Research Foundation of Korea (Grant No. 2011-0008907). 
The authors acknowledge useful discussions with Han Woong Yeom and Tae-Hwan Kim.


\newcommand{\PR} [3]{Phys.\ Rev.\ {\bf #1}, #2 (#3)}
\newcommand{\PRL}[3]{Phys.\ Rev.\ Lett.\ {\bf #1}, #2 (#3)}
\newcommand{\PRB}[3]{Phys.\ Rev.\ B\ {\bf #1}, #2 (#3)}
\newcommand{\PST}[3]{Phys.\ Scr.\ T\ {\bf #1}, #2 (#3)}
\newcommand{\PML}[3]{Phil.\ Mag.\ Lett.\ {\bf #1}, #2 (#3)}
\newcommand{\SCI}[3]{Science {\bf #1}, #2 (#3)}
\newcommand{\SSA}[3]{Surf.\ Sci.\ {\bf #1}, #2 (#3)}
\newcommand{\SSCO}[3]{Solid\ State\ Comm.\ {\bf #1}, #2 (#3)}
\newcommand{\SSL}[3]{Surf.\ Sci.\ Lett.\ {\bf #1}, #2 (#3)}
\newcommand{\SRL}[3]{Surf.\ Rev.\ Lett.\ {\bf #1}, #2 (#3)}
\newcommand{\NAT}[3]{Nature\ Phys.\ {\bf #1}, #2 (#3)}
\newcommand{\JP}[3]{J.\ Phys.\ {\bf #1}, #2 (#3)}
\newcommand{\JACS}[3]{J.\ Am.\ Chem.\ Soc.\ {\bf #1}, #2 (#3)}
\newcommand{\JAP}[3]{J.\ Appl.\ Phys.\ {\bf #1}, #2 (#3)}
\newcommand{\JCP}[3]{J.\ Chem.\ Phys.\ {\bf #1}, #2 (#3)}
\newcommand{\JPCS}[3]{J.\ Phys.\ Chem.\ Solids.\ {\bf #1}, #2 (#3)}
\newcommand{\JVSA}[3]{J.\ Vac.\ Sci.\ Technol.\ A\ {\bf #1}, #2 (#3)}
\newcommand{\JVSB}[3]{J.\ Vac.\ Sci.\ Technol.\ B\ {\bf #1}, #2 (#3)}
\newcommand{\JJAP}[3]{Jpn.\ J.\ Appl.\ Phys.\ {\bf #1}, #2 (#3)}
\newcommand{\ASS}[3]{Appl.\ Surf.\ Sci.\ {\bf #1}, #2 (#3)}
\newcommand{\APL}[3]{Appl.\ Phys.\ Lett.\ {\bf #1}, #2 (#3)}
\newcommand{\CPL}[3]{Chem.\ Phys.\ Lett.\ {\bf #1}, #2 (#3)}
\newcommand{\LTP}[3]{Low\ Temp.\ Phys.\ {\bf #1}, #2 (#3)}
\newcommand{\TSF}[3]{Thin\ Solid\ Filims\ {\bf #1}, #2 (#3)}
\newcommand{\VAC}[3]{Vacuum\ {\bf #1}, #2 (#3)}
\newcommand{\JPCC}[3]{J.\ Phys.\ Chem.\ C\ {\bf #1}, #2 (#3)}
\newcommand{\RPP}[3]{Rep.\ Prog.\ Phys.\ {\bf #1}, #2 (#3)}
\newcommand{\JPCM}[3]{J.\ Phys.\ Condens.\ Matter.\ {\bf #1}, #2 (#3)}
\newcommand{\CJCP}[3]{Chin.\ J.\ Chem.\ Phys.\ {\bf #1}, #2 (#3)}
\newcommand{\IJMPB}[3]{Int.\ J.\ Mod.\ Phys.\ B\ {\bf #1}, #2 (#3)}
\newcommand{\RMP}[3]{Rev.\ Mod.\ Phys.\ {\bf #1}, #2 (#3)}
\newcommand{\NNT}[3]{Nanotechnology\ {\bf #1}, #2 (#3)}
\newcommand{\JPD}[3]{J.\ Phys.\ D\ {\bf #1}, #2 (#3)}
\newcommand{\JPC}[3]{J.\ Phys.\ C\ {\bf #1}, #2 (#3)}
\newcommand{\PSSC}[3]{Phys.\ Status\ Solidi\ C\ {\bf #1}, #2 (#3)}
\newcommand{\EL}[3]{Europhys.\ Lett.\ {\bf #1}, #2 (#3)}

\end{document}